\documentclass[
reprint,
superscriptaddress,
amsmath,amssymb,
aps,
pra,
]{revtex4-1}
\usepackage{float}
\usepackage{xcolor}

\usepackage{graphicx}% Include figure files
\usepackage{dcolumn}% Align table columns on decimal point
\usepackage{bm}% bold math
\usepackage{hyperref}% add hypertext capabilities

\graphicspath{{figures/}}

\begin{document}

\preprint{APS/123-QED}
\title{Comparing tunneling spectroscopy and charge sensing of \\ Andreev bound states in a semiconductor-superconductor hybrid nanowire structure}

\author{Deividas~Sabonis}
 \thanks{These authors contributed equally to this work}

\affiliation{Center for Quantum Devices, Niels Bohr Institute, University of Copenhagen, 2100 Copenhagen, Denmark}
\affiliation{Microsoft Quantum Lab--Copenhagen, Niels Bohr Institute, University of Copenhagen, 2100 Copenhagen, Denmark}
 
\author{David~van~Zanten}
\thanks{These authors contributed equally to this work}

\affiliation{Center for Quantum Devices, Niels Bohr Institute, University of Copenhagen, 2100 Copenhagen, Denmark}
\affiliation{Microsoft Quantum Lab--Copenhagen, Niels Bohr Institute, University of Copenhagen, 2100 Copenhagen, Denmark}

\author{Judith~Suter}
\affiliation{Center for Quantum Devices, Niels Bohr Institute, University of Copenhagen, 2100 Copenhagen, Denmark}
\affiliation{Microsoft Quantum Lab--Copenhagen, Niels Bohr Institute, University of Copenhagen, 2100 Copenhagen, Denmark}

 \author{Torsten~Karzig}
\affiliation{Microsoft Quantum, Station Q, University of California, Santa Barbara, California 93106-6105, USA}

\author{Dmitry~I.~Pikulin}
\affiliation{Microsoft Quantum, Station Q, University of California, Santa Barbara, California 93106-6105, USA}
\affiliation{Microsoft Quantum, Redmond, Washington 98052, USA}

 \author{Jukka~I.~V\"{a}yrynen}
 \affiliation{Microsoft Quantum, Station Q, University of California, Santa Barbara, California 93106-6105, USA}
\affiliation{Department of Physics and Astronomy, Purdue University, West Lafayette, Indiana 47907, USA}

 \author{Eoin~O'Farrell}
\affiliation{Center for Quantum Devices, Niels Bohr Institute, University of Copenhagen, 2100 Copenhagen, Denmark}
\affiliation{Microsoft Quantum Lab--Copenhagen, Niels Bohr Institute, University of Copenhagen, 2100 Copenhagen, Denmark}

 \author{Davydas~Razmadze}
\affiliation{Center for Quantum Devices, Niels Bohr Institute, University of Copenhagen, 2100 Copenhagen, Denmark}
\affiliation{Microsoft Quantum Lab--Copenhagen, Niels Bohr Institute, University of Copenhagen, 2100 Copenhagen, Denmark}

 \author{Peter~Krogstrup}
\affiliation{Center for Quantum Devices, Niels Bohr Institute, University of Copenhagen, 2100 Copenhagen, Denmark}
\affiliation{Microsoft Quantum Materials Lab--Copenhagen, 2800 Lyngby, Denmark}

\author{Charles~M.~Marcus}
\affiliation{Center for Quantum Devices, Niels Bohr Institute, University of Copenhagen, 2100 Copenhagen, Denmark}
\affiliation{Microsoft Quantum Lab--Copenhagen, Niels Bohr Institute, University of Copenhagen, 2100 Copenhagen, Denmark}


\begin{abstract}

Transport studies of Andreev bound states (ABSs) are complicated by the interplay of charging effects and superconductivity. Here, we compare transport approaches to ABS spectroscopy in a semiconductor-superconductor island to a charge-sensing approach based on an integrated radio-frequency single-electron transistor. Consistency of the methods demonstrates that fast, non-invasive charge sensing allows accurate quantitative measurement of ABSs while eluding some complexities of transport.
  
\end{abstract}

	\maketitle

%\tableofcontents
 Experiments with superconductor-semiconductor hybrid structures have demonstrated the presence of Andreev bound states (ABSs) in proximitized semiconductors by a variety of methods, including tunneling spectroscopy in NIS junctions (N is normal, S is superconductor, I is an insulating tunnel barrier), Coulomb blockade spectroscopy in superconducting islands with normal leads (NISIN) \cite{paritylifetime,albrecht2016exponential, shen2018parity, deng2016majorana,vaitiekenas2020flux}, and SNS spectroscopy of Josephson junctions \cite{laroche2019observation,van2019photon,sabonis2019,kringhoj2020andreev, tosi2019spin,hays2020continuous,hays2018direct, van2017microwave}.  These methods reveal distinct spectroscopic features and raise different technical challenges.  With  growing interest in detecting and controlling Majorana zero modes (MZMs) across Josephson junctions \cite{ginossar2014microwave,hassler2011top,hyart2013flux} and distinguishing them from ABSs, it is important to compare  methods in systems that can be probed several ways. 
 
\begin{figure}
\centering
\includegraphics[width=0.42 \textwidth]{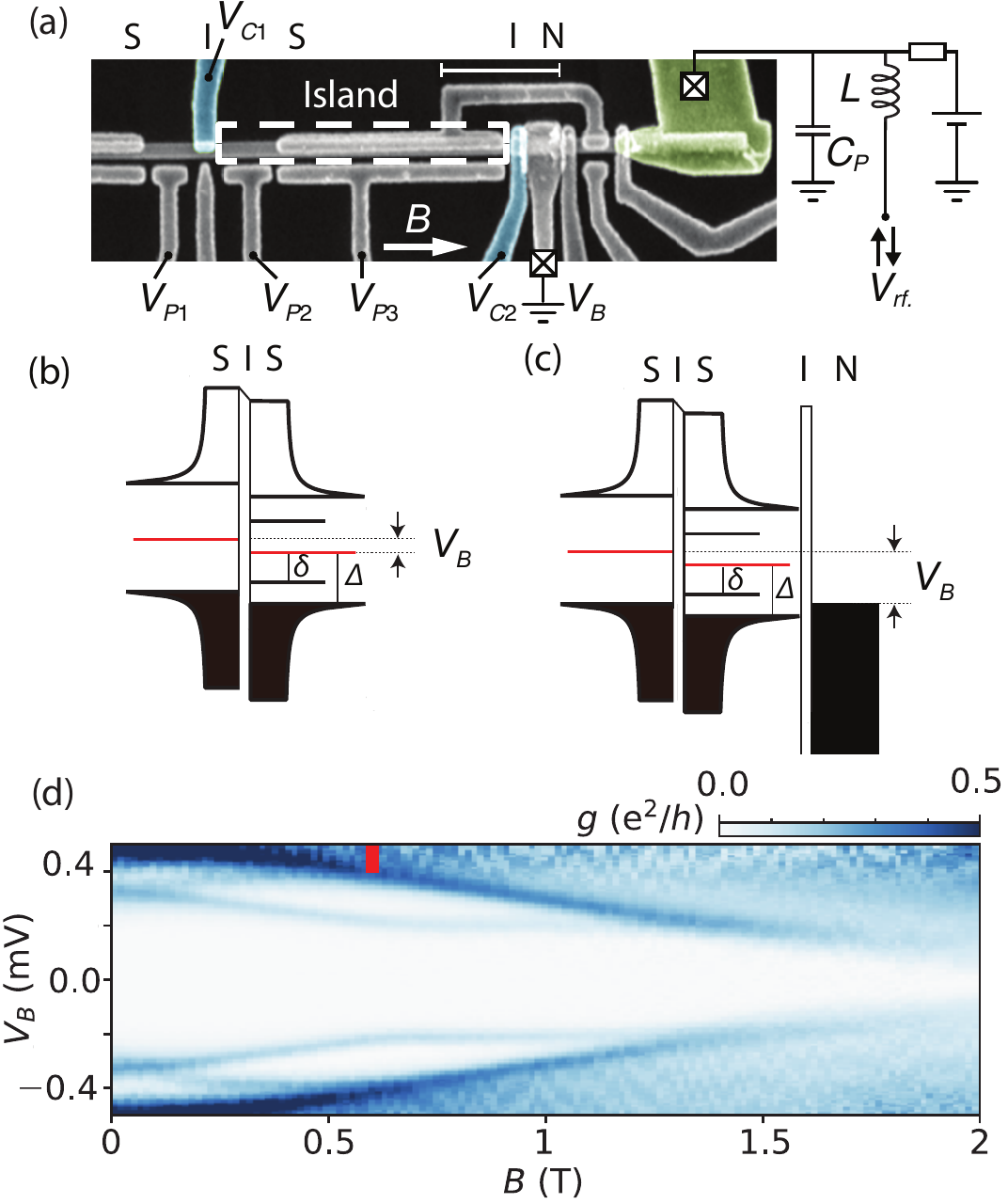}
\caption{(a) False coloured scanning electron micrograph of the device consisting of a InAs nanowire with segments of epitaxial Al (scale bar 1~$\mu$m). The main segment of the nanowire defines the Coulomb blockaded superconducting island (S) that is separated with the gate-tuneable barriers (I) from the superconducting segment on the left (S) and normal segment (N) on the right. Voltages applied to plunger gates P1, P2 and P3 tune the carrier density in the proximitized regions whereas gates C1 and C2 are used for controlling the tunnel barrier height. The island is capacitively coupled to a single-electron transistor charge sensor at the right end of the nanowire that is a part of the resonant circuit connected to a right lead (green). (b) Schematics of the superconductor-insulator-superconductor system (SIS). The induced superconducting gap ($\Delta$) and subgap state energy ($\delta$) on the island side are indicated together with bias voltage $V_B$. (c) Schematics of the superconducting island system confined by barriers (I) from superconducting contact on the left (S) and normal contact on the right (N). (d) SIS tunneling conductance $g$ as a function of axial magnetic field $B$ and source-drain bias $V_{B}$. Starting at $B$~=~0.6~T (red marker) the feature in conductance at $V_{B}$~=~220~$\mu$V becomes independent on the further increase in magnetic field.}
\label{fig1}
\end{figure}

In this Article, we report measurements of an electrostatically gated multi-segment hybrid InAs/Al nanowire (NW) device [see Fig.~\ref{fig1}(a)] that is reconfigurable {\it in situ}, allowing a comparison of spectroscopic signatures of the {\it same} ABS by three methods: a Josephson junction (SIS), a Coulomb island (SISIN), and a radio-frequency (RF) charge sensor. We observed consistency between methods, noting that RF charge detection was especially fast, and, unlike transport, does not alter charge occupancy during measurement.  
 
 The device was fabricated from an InAs NW grown by molecular beam epitaxy, with Al grown on three facets of the hexagonal core. Using conventional lithographic processing, Al was removed by wet etching in $\sim$100~nm segments adjacent to cutter gates C1 and C2, providing gate-controllable barriers. Blanket atomic layer deposited ${\rm HfO}_2$ insulated all gates from the NW. Depleting the InAs with C1 and C2 yielded an SISIN device with an S island separated by  tunnel barriers (I) from an S lead on the left and N lead on the right. Voltages applied to plunger gates P1-P3 tuned carrier density in adjacent proximitized semiconductor regions as well as the charge offset on the S island. At the far right, a bare InAs segment (again, created by etching of Al), served as RF single-electron transistor (RF-SET) that was capacitively coupled to the main S island by a floating gate [see Fig.~\ref{fig1}(a)]. Data acquisition followed Ref.~\cite{van2019photon} and charge sensing followed Ref.~\cite{Razmadze2018}. Measurements were performed in a dilution refrigerator with a 6-1-1~T vector magnet.

To generate ABSs in the nanowire, we applied a moderate magnetic field, $B$~=~600 mT, along the nanowire axis and set gate C1 to the tunneling regime (conductance $g \ll e^{2}/h$), junction C2 to the open regime ($g \sim e^2 /h$), and gates P1-P3 to $-4$~V to reduce density in the NW. In this configuration, SIS spectroscopy [Figs.~\ref{fig1}b)] was used to locate a single subgap state by tuning P3. Figure~\ref{fig1}(d) shows differential conductance $g$ as a function of bias $V_{B}$ and $B$ in this SIS configuration. Near $B=0$ a precursor of the finite-field ABS is seen at $V_{B}~\sim~\pm330~\mu$V. Increasing $B$ leads to a single ABS moving down with an effective $g$ factor of $\sim 6$. Starting at $B\sim 0.6$~T [red marker in Fig.~\ref{fig1}(d)] the feature at $V_{B}$~=~220~$\mu$V becomes independent of the further increase $B$. This presumably results from competition between a reduction of the induced superconducting gap and a zero-energy-crossing ABS \cite{shen2018parity}. Unless otherwise noted, measurements described below are at $B$~=~0.6~T. 

\begin{figure}
    \includegraphics[width=0.42\textwidth]{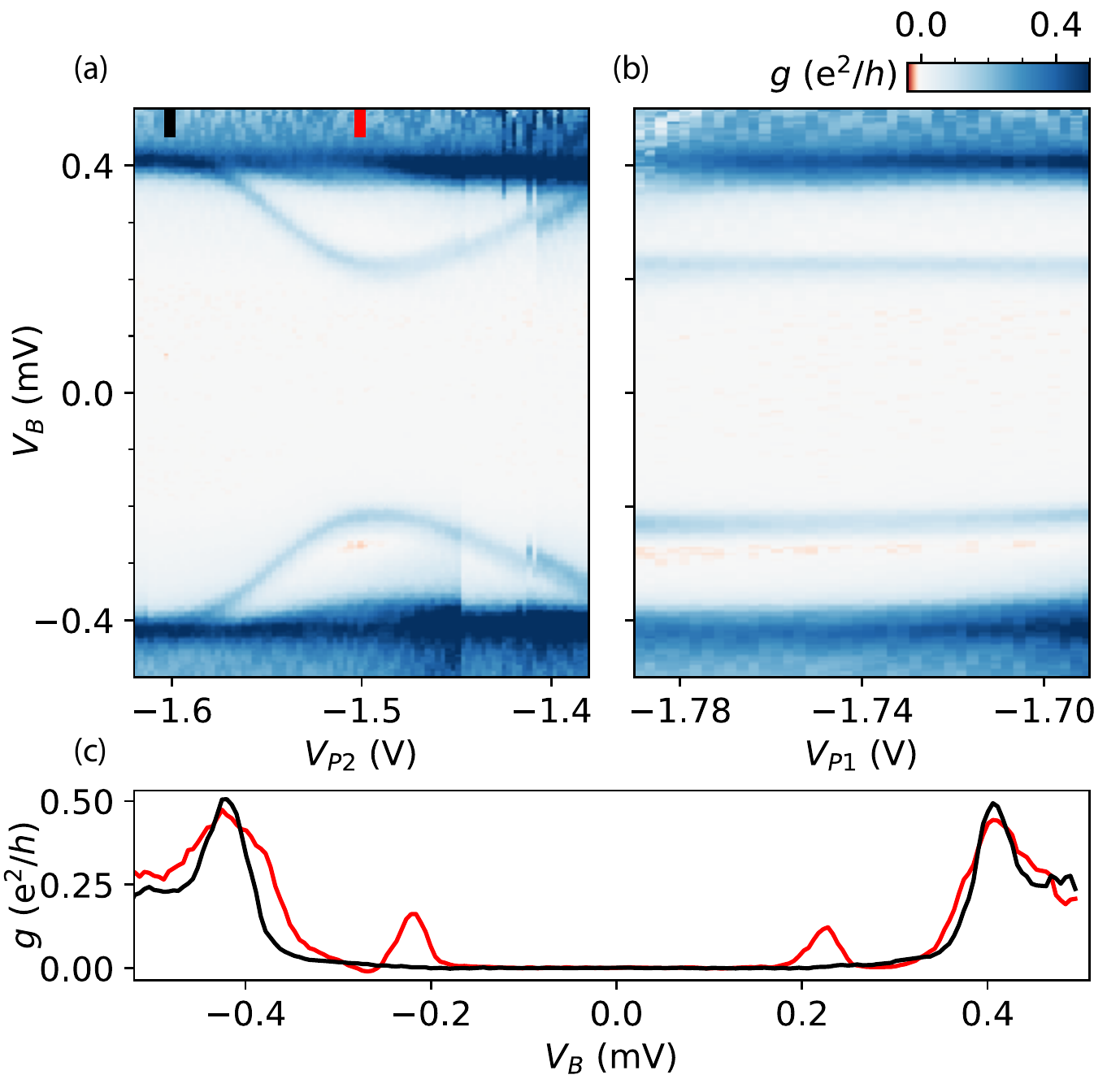}
    \centering
    \caption{(a) Two dimensional map of superconductor-insulator-superconductor tunneling differential conductance $g$, as a function of source-drain bias $V_{B}$ and gate voltage $V_{P2}$ at $B$~=~0.6~T. ABS shows a response to the voltage on plunger P2. (b) Differential conductance $g$ as a function of $V_{B}$ and voltage $V_{P1}$ measured at $V_{P2}$~=~$-$1.490~V. A weak response of the ABS with respect to gate P1 indicates the location of the state as being on the right side of the barrier (superconducting island). (c) Line-cuts at red and black marker positions in (a), at the minimum ABS energy (red) and with the ABS energy merging with the continuum (black).}
    \label{fig2}
\end{figure}

The location of the ABS within the NW could be inferred from its dependence on P1 and P2. Specifically, spectroscopy showed little response to $V_{P1}$ and a strong response to $V_{P2}$, as shown in Figs.~\ref{fig2}(a,b), indicating that the ABS was located on the right side of cutter gate C1, as represented in Fig.~\ref{fig1}(b). The energy of the ABS approached the gap edge for $V_{P2}\sim -1.6$~V and $\sim -1.35$~V with a minimum at around $V_{P2}\sim -1.5$~V. Differential conductance as a function of bias, $V_{B}$, and gate P1 [Fig.~\ref{fig2}(c)] were carried out at $V_{P2}=-1.49$~V.

\begin{figure}
\centering
    \includegraphics[width=0.38\textwidth]{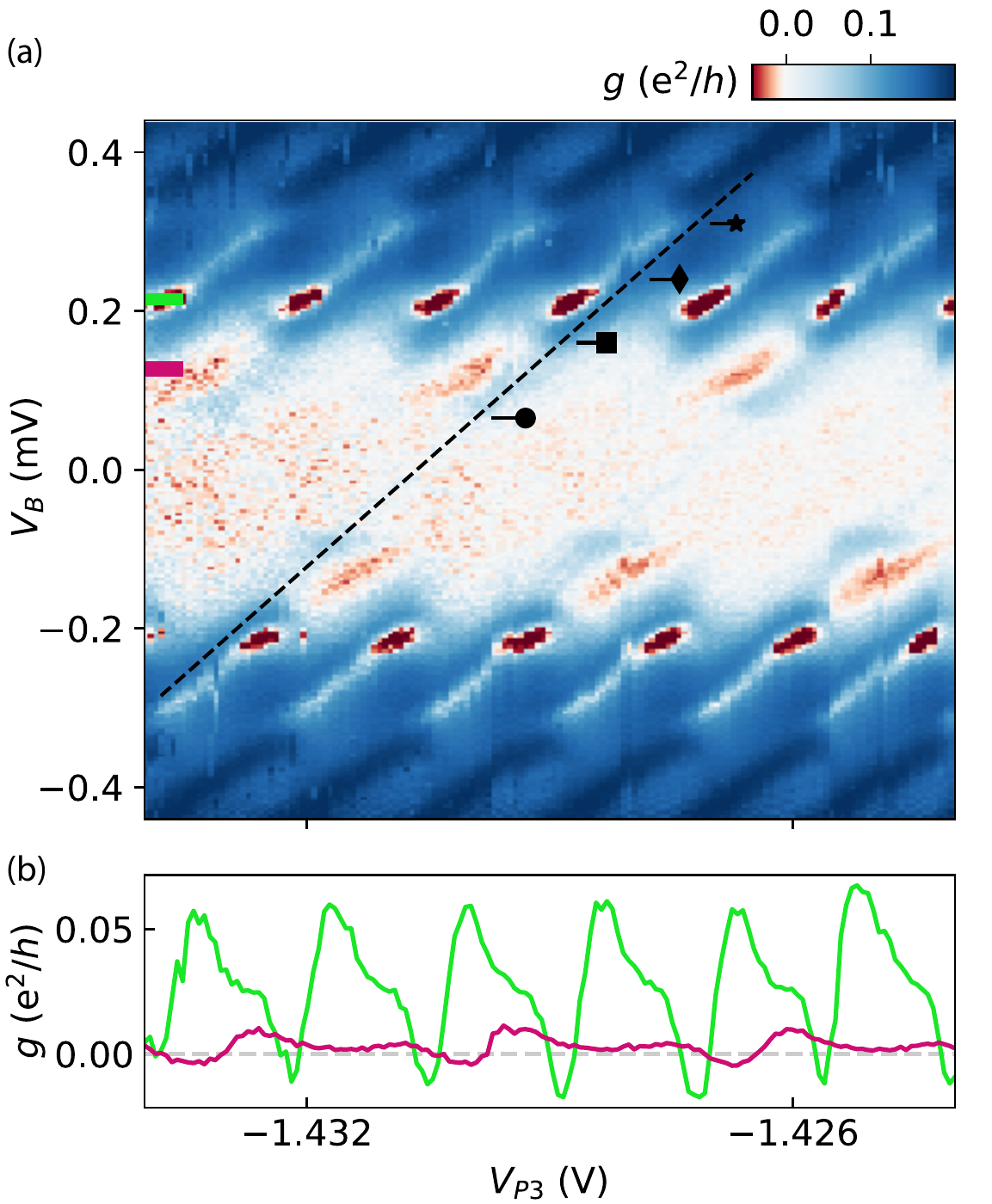}
    \caption{(a) Coulomb blockade measured in differential conductance $g$ as a function of the island plunger voltage $V_{P3}$ and a source-drain bias $V_{B}$ at $V_{P2}$~=~$-$1.456~V and axial magnetic field $B$~=~0.6~T. Pink marker indicates the bias $V_B$ value at which the 2$e$-periodic negative differential conductance is observed whereas green marker indicates the position of 1$e$ negative differential conductance. The diagonal dashed line indicates the $V_B$ - $V_{P3}$ configuration for which the superconducting island and the left superconducting lead stay at the same potential. Four markers along the diagonal trace indicated the threshold values for the bias that enable the transport processes discussed in the main text. (b) Line-cuts from (a) at the pink and the green marker positions in (a) indicating 2$e$ and 1$e$-periodic features together with the associated negative differential conductance.}
    \label{fig3}
\end{figure}

Following SIS spectroscopy,  gates C1 and C2 were set to the tunneling regime to create a superconducting island, providing SISIN spectroscopy. Figure~\ref{fig3}(a) shows the resulting two-dimensional conductance map as a function of bias and island plunger voltage $V_{P3}$. In the SISIN configuration, transport showed 1$e$-periodic resonances at high bias, $V_{B}~\gtrsim~400~\mu$V, and 2$e$-periodic features at low bias. The black dashed diagonal line in Fig.~\ref{fig3}(a) indicates the alignment of the chemical potential of the left lead and the island. Four features along this diagonal are highlighted: Near $V_B = 0$, faint 2$e$-periodic features that become stronger along the diagonal at $V_B \sim 65~\mu$V ($\bullet$), accompanied by negative differential conductance (NDC); Further increasing bias yielded stronger 1$e$-periodic features starting at $V_B \sim 160~\mu$V ($\blacksquare$) along with stronger NDC; Around $V_B \sim 240~\mu$V ($\blacklozenge$) the conductance increase was largely independent of $V_{P3}$; Finally, at $V_B~\sim~300~\mu$V the strong 1$e$-periodic conductance resonances ($\bigstar$) were observed.

These marked features in Fig.~\ref{fig3}(a) are consistent with a model of an SISIN device with a single ABS at energy $\delta$ on the S island with electrostatic energy $E_C(n-n_g)^2$, where $E_C$ is the charging energy, $n$ is the number of electrons on the island, and $n_g$ is the dimensionless gate voltage.  In particular, the 2$e$-periodic features at zero bias arise from resonant Cooper pair tunneling between S lead and S island at odd integer values of $n_g$. At increased bias, the resonance occurs at $2V_B = 4E_C n_g$ [dashed line in Fig.~\ref{fig3}(a)], with the S island on resonance with the left S lead while the energy difference from the Fermi level in the N lead increases. For the case $\delta<E_C$, the 2$e$-periodic features are expected to be suppressed around zero bias, as  Cooper pair resonances are between excited states above the odd parity ground state of the occupied ABS close to odd integer values of $n_g$. The 2$e$-periodic features around zero bias in Fig.~\ref{fig3}(a) suggest that indeed $\delta<E_C$. At $V_B = E_C - \delta$ [$\bullet$ in Fig.~\ref{fig3}(a)], the ABS can empty into the right normal lead, enabling resonant Cooper pair tunneling, brightening the 2$e$-periodic feature. At yet larger bias with fixed $n_g$, the resonance condition for  Cooper pair tunneling from left is exceeded, resulting in NDC. At $V_B~=~\Delta_L+\delta-E_C$ [$\blacksquare$ in Fig.~\ref{fig3}(a)] above-gap quasiparticles in the left S lead can excite the ABS thus allowing single-electron transport that no longer requires resonant Cooper pair processes, resulting in 1$e$-periodic resonances in $n_g$. Here, NDC presumably results when the coherence peak in the density of states the S lead surpasses the bound state energy. At $V_B~=~\Delta_L$ [$\blacklozenge$ in Fig.~\ref{fig3}(a)] quasiparticles from the S lead can transfer to the N lead via elastic cotunneling through the ABS or other subgap states, which are only weakly dependent on $n_g$. Finally, at $V_B = \Delta_L+\Delta_I-E_C$ [$\bigstar$ in Fig.~\ref{fig3}(a)] single-electron transport between continuum states of the left S lead and the S island become energetically allowed. Figure ~\ref{fig3}(b) shows line cuts from Fig.~\ref{fig3}(a),  indicating both 2$e$ (pink) and 1$e$ (green) features along with the associated NDC. Further details of the four features are discussed in Appendix \ref{app:theory}.

We extract a charging energy $E_{C}$~$\sim$~85~$\mu$eV from independent transport data of Coulomb diamonds (not shown). Using this value, the above model is consistent with an ABS energy $\delta=20~\mu$eV and superconducting gaps $\Delta_L \sim 220~\mu$eV and $\Delta_I \sim 160~\mu$eV, in the left S lead and S island, both consistent with values seen in similar hybrid nanowires \cite{ChangNNano}. The condition $E_{C}<\Delta_{I}$ is consistent with the  2$e$ features at low bias in Fig.~\ref{fig3}(a). 

Transport features observed in SISIN spectroscopy are rather complicated, even with the relatively simple situation of a single ABS on the island. Previous work has  showed that RF charge readout offers some advantages compared to transport, particularly speed and simplicity of the data \cite{Razmadze2018,van2019revealing,de2019rapid}. To perform charge readout on the present device, we use the integrated charge sensor and examine the same ABS investigated above via transport.

\begin{figure}
\centering
    \includegraphics[width=0.42\textwidth]{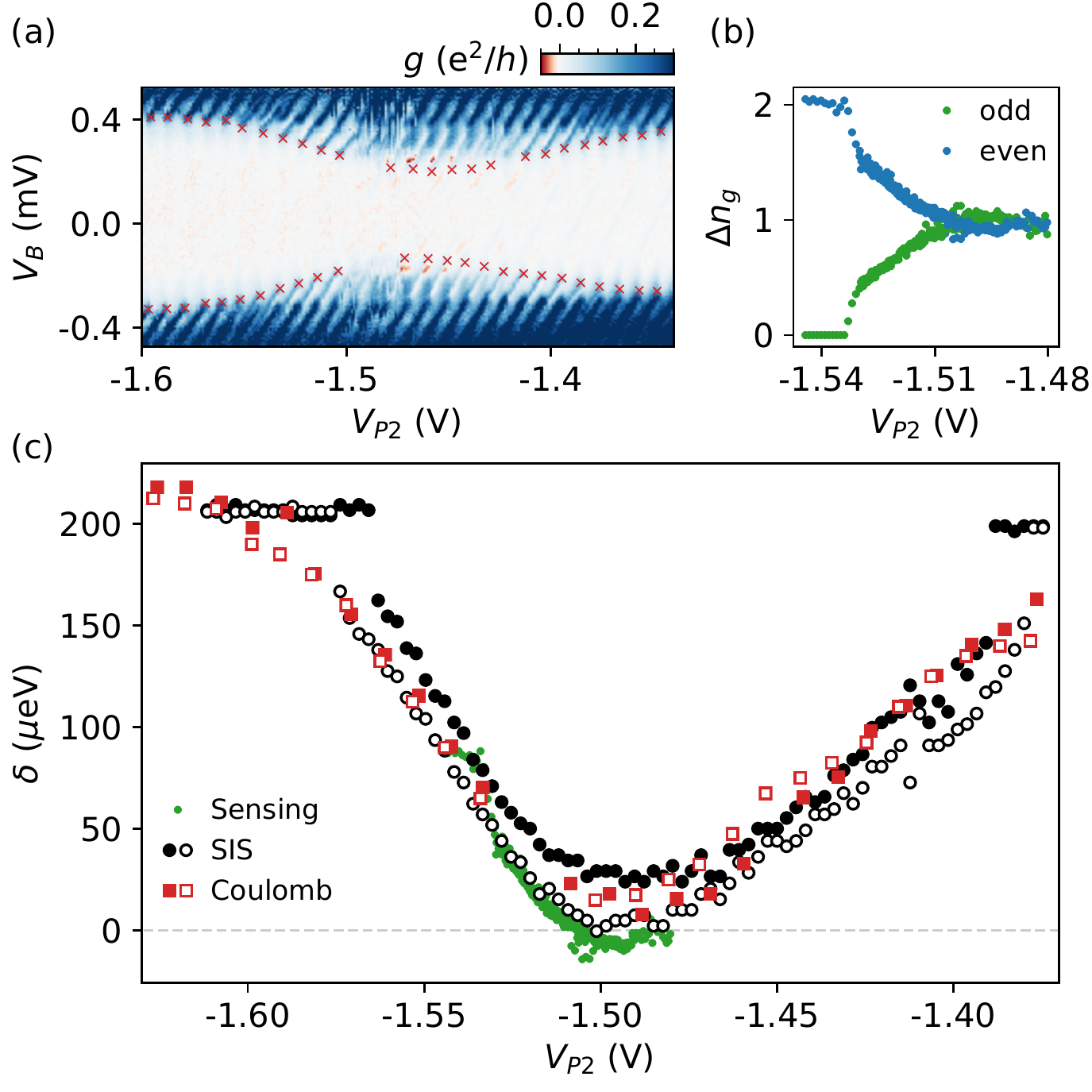}
    \caption{(a) Coulomb blockade spectroscopy measured in conductance $g$ through the island in the extended plunger P2 voltage range. The effect of the ABS can be indicated as an envelope modulation of Coulomb blockade features following the shape of the ABS from the Fig.~\ref{fig2}(a) (red markers around the tip of Coulomb diamonds). (b) Normalised peak spacing of even (blue) and odd (green) Coulomb valleys as a function of $V_{P2}$ extracted from charging sensing data (see Fig.~\ref{fig6}). The 2$e$ to 1$e$ transition in periodicity of Coulomb blockade results from the ABS energy becoming smaller than the charging energy, controlled by $V_{P2}$. (c) Bound state profile and energy extracted by three different methods: SIS transport spectroscopy (black), SISIN Coulomb blockade spectroscopy (red) and RF charge sensing - in a gate voltage regime where the charge sensing data was acquired (green). Green and red data have been shifted horizontally by 30~mV to compensate capacitive coupling to $V_{C2}$.}
    \label{fig4}
\end{figure}

Figure~\ref{fig4}(a) shows transport-based Coulomb blockade spectroscopy in the SISIN configuration performed over a larger range of P2 gate voltages. The effect of the ABS can be seen as an envelope modulation of Coulomb blockade that follows the shape of the ABS that was previously shown in the Fig.~\ref{fig2}(a). To emphasize that the modulation is connected to the ABS, red markers in Fig.~\ref{fig4}(a) indicate the tips of Coulomb diamonds. Charge sensing data from the same gate regime as Fig.~\ref{fig4}(a) is shown in Fig.~\ref{fig6}. Charge sensing shows explicitly the transition in periodicity of superconducting island occupancy whenever the bound state energy $\delta$ is lowered with respect to charging energy $E_C$ of the island. This change in occupancy was implied by the model, but not explicitly visible in transport data. Figure~\ref{fig4}(b) shows the effect of the 2$e$ to 1$e$ transition as a function of plunger voltage $V_{P2}$ measured with the the integrated RF-SET. The small overshoot of the even level (blue) at $V_{P2}$~=~$-$1.51~V is associated to ABS that is oscillating around zero energy as a function of $V_{P2}$. Gate P3 in the Fig.~\ref{fig6} is also responsible for tuning the ABS however with a smaller lever arm compared to P2 compatible with the ABS being localized closer to the barrier C1 side. We note that the suppression of the 2$e$-periodic signal around zero bias in Fig.~\ref{fig4}(a) is correlated with the onset of 1$e$-periodicity in Fig.~\ref{fig4}(b). This indicates that the exchange of Cooper pairs between the island and superconducting left lead, despite being aligned in chemical potential, is indeed suppressed when $\delta<E_C$. The latter requirement is supported by the observation of re-entrant resonant Cooper pair processes at $V_B~=~E_C-\delta$.

The three measurements methods of the same ABS are compared in Fig.~\ref{fig4}(c). Black data points show the ABS profile extracted using SIS spectroscopy [Fig.~\ref{fig2}(a)], red - Coulomb blockade measurements where the ABS energy $\delta$ is determined using the onset of Coulomb blockade features [Fig.~\ref{fig4}(a)] and green - RF-SET measurements [Fig.~\ref{fig6}]. For SIS and SISIN transport measurements both positive and negative values of $V_B$ were used. An offset in $V_{C2}$ of $\sim$~30~mV between SIS and SISIN measurements, presumably due to cross-capacitance to $V_{C2}$ has been removed. Correspondence between the ABS energy profiles measured using transport (SIS and SISIN) and charge sensing shows that the charge periodicity of the island is indeed affected by the ABS energy. It also shows that RF measurements with increased acquisition rate \cite{Razmadze2018,de2021rapid} is a useful alternative to  transport.

\section*{Appendices}
\subsection{Transport processes}
\begin{figure}
\centering
    \includegraphics[width=0.3\textwidth]{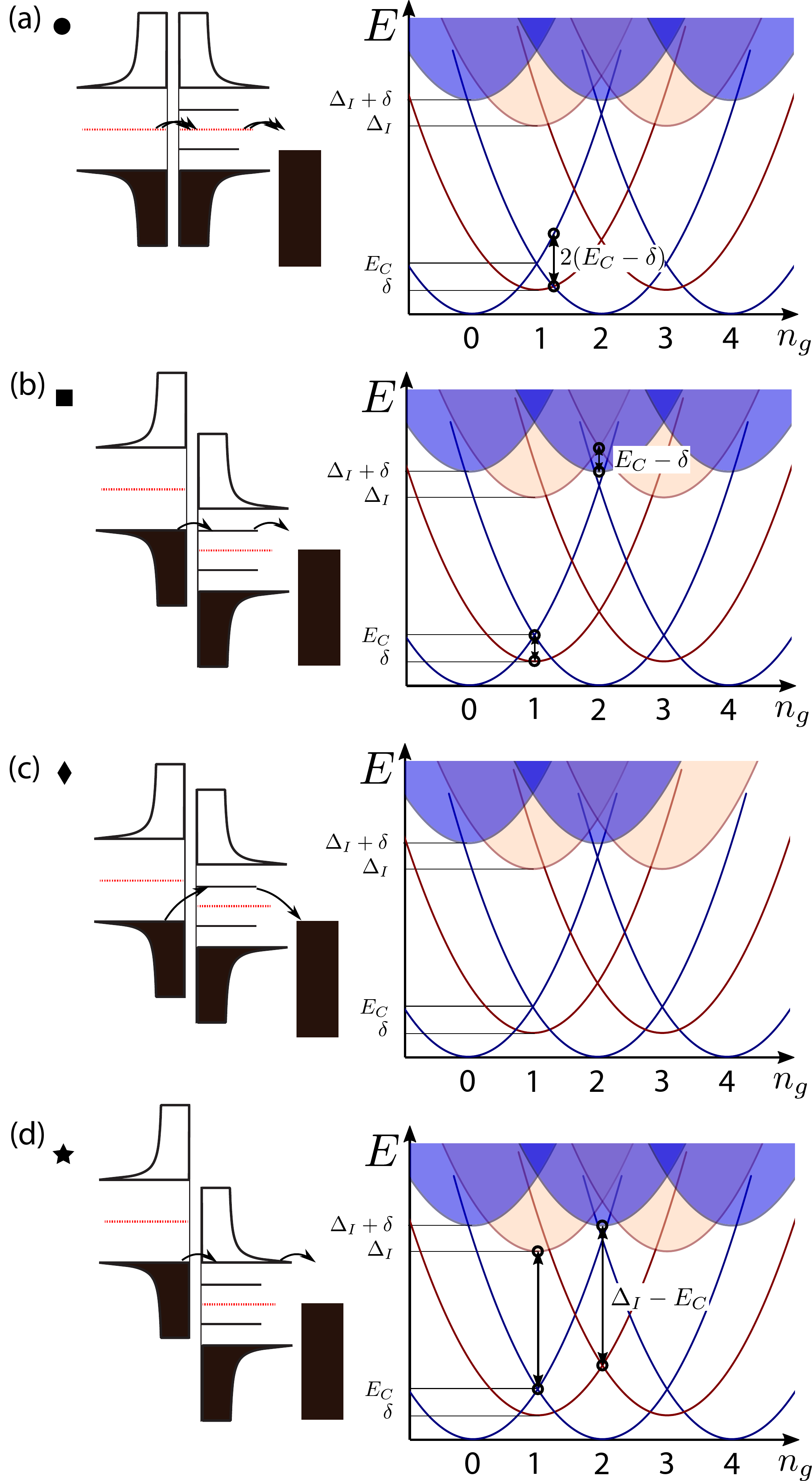}
    \caption{Model of transport processes through the superconducting island for different bias configurations. (a) 2$e$-periodic process at $V_B$~=~$E_C$ $-$ $\delta$. (b) 1$e$-periodic process at bias $V_B$~=~$\Delta_L$ $-$ ($E_C$ $-$ $\delta$). (c) 1$e$-periodic process constant in $n_g$ at $V_B$~=~$\Delta_L$ (elastic cotunneling through the bound state or other subgap states). (d) 1$e$-periodic process at bias $V_B$~=~$\Delta_L$~+~$\Delta_I$~$-$~$E_C$. This process is similar to (b). The system can save an energy $E_C$ for the threshold when going from even to odd parity state around odd integer $n_g$ values or from odd to even parity at even integer $n_g$.}
    \label{fig5}
\end{figure}

Figure~\ref{fig5} shows schematics of the transport processes corresponding to the markers in Fig.~\ref{fig3}(a), within our model. In Fig.~\ref{fig5}(a) the 2$e$-periodic process at $V_B$~=~$E_C$ $-$ $\delta$ ($\bullet$) is visualized. Typically Cooper pair transfer is resonant around odd integer $n_g$ values. However, in the case of a subgap state energy $\delta < E_C$ the 2$e$ transport is blocked. At $V_B > E_C$ $-$ $\delta$ the transport can get unblocked by emptying the subgap state into the right normal lead. Panel (b) depicts 1$e$-periodic process at bias $V_B$~=~$\Delta_L$ $-$ ($E_C$ $-$ $\delta$) ($\blacksquare$). The island can relax its energy when going from the even to the odd parity state around odd integer $n_g$ values or from odd parity to even parity at even $n_g$ values. The relaxation reduces the bias $V_B$ threshold for single-electron transfers from the left lead. In Fig.~\ref{fig5}(c) a single-electron process at $V_B$~=~$\Delta_L$ ($\blacklozenge$) can be associated with elastic cotunneling through the bound state or other subgap states. In Fig.~\ref{fig5}(d), again, a 1$e$-periodic process at bias $V_B$~=~$\Delta_L$~+~$\Delta_I$~$-$~$E_C$ ($\bigstar$) happens. This process is similar to (b). The system can save an energy $E_C$ for the threshold when going from even parity state to odd parity state around odd integer $n_g$ or from odd to even parity at even integer values of $n_g$. Both (b) and (d) require poisoning or some other way of bringing the island out of its ground state. Since both of these are at substantial bias $V_B$ $\sim$ $\Delta_L$ it becomes possible to create the corresponding excited quasiparticle states, for example by splitting a Cooper pair from the left lead (thus gaining energy $2V_B$) to create two quasiparticles on the island.

\subsection{Charge sensing of bound state}
\label{app:theory}

\begin{figure}
\centering
    \includegraphics[width=0.44\textwidth]{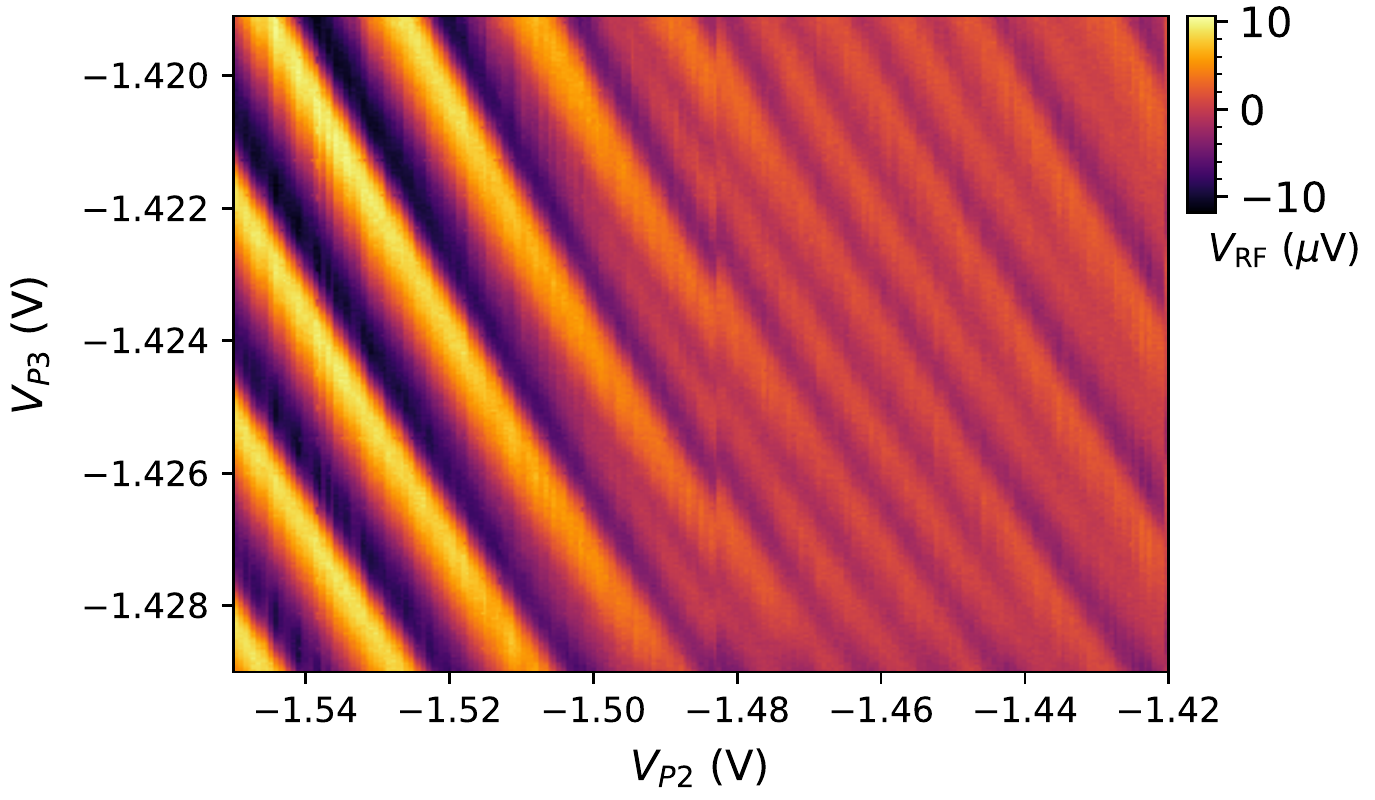}
    \caption{Two dimensional gate map measured in charge sensor response $V_{\rm RF}$ as a function of plunger voltages $V_{P3}$ and $V_{P2}$. The map shows that the island changes its periodicity from the 2$e$ at lower end of $V_{P2}$ voltages, to even-odd and eventually 1$e$ whenever the energy of the bound state $\delta$ is lowered below the charging energy $E_{C}$ of the superconducting island. The system recovers 2$e$ periodicity for more positive values of $V_{P2}$. 
    }
    \label{fig6}
\end{figure}

The charge sensor was capacitively coupled to the superconducting island via floating coupler. Fig.~\ref{fig6} shows a two dimensional gate map recorded in demodulated voltage $V_{\rm RF}$ from the charge sensor as a function of gate voltages $V_{P2}$ and $V_{P3}$ (line average along $V_{P3}$ axis subtracted). Measurements are performed at $V_{B}$~=~0. Constant charge positions appear as bright plateaus interrupted by transitions between the two charge states (dark). The jump at $V_{P2}$~=~$-$1.485~V is likely due to electrostatic background charges in the nanowire environment or in the dielectric covering the nanowire. The effect of the ABS can be seen as a change of gate space periodicity along the $V_{P2}$ axis. It starts from 2$e$-periodic behaviour at more negative values of $V_{P2}$. Then as the ABS energy $\delta$ is lowered starting at $V_{P2}$~=~$-$1.490~V it transitions through the even-odd and eventually 1$e$ regime recovering back to 2$e$ for more positive values of $V_{P2}$. Gate P3 is also responsible for tuning the ABS however with a smaller lever arm compared to P2 compatible with the ABS being localised closer to the barrier C1 side.

\subsection{Fabrication and measurement}

The 100~nm diameter nanowire is grown using the vapor-liquid-solid technique in a molecular beam epitaxy system with the InAs [111] substrate crystal orientation. Following the growth, Al is deposited epitaxially \textit{in situ} on three facets of the nanowire with an average thickness of 10~nm. The nanowire is then positioned by hand on the substrate using a micro-manipulator tool, which gives few-micrometer placement precision. Using standard lithography, the nanowire is patterned and Al is etched away where tunneling barrier controlling gates are defined. Ti/Au contacts and gates are then patterned and thermally evaporated. 

Signal demodulation was carried out using a Zurich Instruments Ultra High Frequency Lock-in (ZIUHFLI) \cite{zi} which uses digital processing but the principle of operation is similar to analog demodulation. 

\section*{Acknowledgments}

We thank Shivendra~Upadhyay for help with fabrication. Research was supported by Microsoft, the Danish National Research Foundation, and the European Research Council under grant 716655. Judith Suter acknowledges financial support from the Werner Siemens Foundation Switzerland.  

% \nocite{*}
% \bibliographystyle{apacite}
\bibliography{Boundstate6.4}

\end{document}